\documentclass[prl,twocolumn,superscriptaddress,preprintnumbers,nofootinbib]{revtex4}
\usepackage{graphicx}
\usepackage{epsfig}
\usepackage{bm}
\usepackage{latexsym,amsmath,amssymb,wasysym,float}
\usepackage{mathrsfs}  
\usepackage{color}
\usepackage{tablefootnote}
\usepackage{enumitem}

\newcommand{\postscript}[2]{\setlength{\epsfxsize}{#2\hsize}
   \centerline{\epsfbox{#1}}}

\usepackage{ dsfont }
\usepackage{float}
\usepackage{enumitem}
\usepackage{mathrsfs}  
\usepackage{hyperref}
\usepackage{multirow}

\usepackage[usenames,dvipsnames]{xcolor}
\definecolor{orange}{cmyk}{0,0.5,1,0}
\definecolor{rossoCP3}{cmyk}{0,.88,.77,.40}
\definecolor{graa}{rgb}{0.8,0.8,0.8}
\definecolor{blaa}{rgb}{0.2,0.2,0.6}

\def\simlt{\mathrel{\lower2.5pt\vbox{\lineskip=0pt\baselineskip=0pt
           \hbox{$<$}\hbox{$\sim$}}}}
\def\simgt{\mathrel{\lower2.5pt\vbox{\lineskip=0pt\baselineskip=0pt
           \hbox{$>$}\hbox{$\sim$}}}}

\newcommand{\be}{\begin{equation}}
	\newcommand{\ee}{\end{equation}}
\newcommand{\ba}{\begin{eqnarray}}
	\newcommand{\ea}{\end{eqnarray}}

 \allowdisplaybreaks

\newcommand{\nua}[1]{\ensuremath{\rlap
           {\kern-2.5pt\ensuremath
           {\overset{\scriptscriptstyle(-)}{\phantom{\nu}}}}
           {\ensuremath{{\nu}_{#1}}}}}

\begin{document}

\preprint{MPP-2023-169}
\preprint{LMU-ASC 29/23}

\title{\color{rossoCP3} Searching for neutrino-modulino oscillations at the Forward
  Physics Facility}

\author{\bf Luis A. Anchordoqui}

\affiliation{Department of Physics and Astronomy,  Lehman College, City University of
  New York, NY 10468, USA
}

\affiliation{Department of Physics,
 Graduate Center, City University
  of New York,  NY 10016, USA
}

\affiliation{Department of Astrophysics,
 American Museum of Natural History, NY
 10024, USA
}

\author{\bf Ignatios Antoniadis}

\affiliation{Laboratoire de Physique Th\'eorique et Hautes \'Energies - LPTHE, Sorbonne Universit\'e, CNRS, 4 Place Jussieu, 75005 Paris, France
}

\affiliation{Department of Mathematical Sciences, University of Liverpool
Liverpool L69 7ZL, United Kingdom}

\author{\bf Karim Benakli}

\affiliation{Laboratoire de Physique Th\'eorique et Hautes \'Energies - LPTHE, Sorbonne Universit\'e, CNRS, 4 Place Jussieu, 75005 Paris, France
}

\author{\bf Jules Cunat}

\affiliation{Laboratoire de Physique Th\'eorique et Hautes \'Energies - LPTHE, Sorbonne Universit\'e, CNRS, 4 Place Jussieu, 75005 Paris, France
}

\author{\bf Dieter L\"ust}

\affiliation{Max--Planck--Institut f\"ur Physik,  
 Werner--Heisenberg--Institut,
80805 M\"unchen, Germany
}

\affiliation{Arnold Sommerfeld Center for Theoretical Physics, \\
Ludwig-Maximilians-Universit\"at M\"unchen,
80333 M\"unchen, Germany
}

\begin{abstract}
  \vskip 2mm \noindent We make use of swampland conjectures to explore
  the phenomenology of neutrino-modulino mixing in regions of the
  parameter space that are within the sensitivity of experiments at
  the CERN's Forward Physics Facility (FPF). We adopt the working assumption
  of Dirac mass terms which couple left- and right-handed
  neutrinos. We further assume that the 3 right-handed neutrinos are
  0-modes of bulk 5-dimensional states in the dark dimension, a novel
  scenario which has a compact space with characteristic length-scale
  in the micron range that produces a natural suppression of the
  4-dimensional Yukawa couplings, yielding naturally light Dirac
  neutrinos. We formulate a specific realization of models with high-scale
  supersymmetry breaking that can host a rather heavy gravitino ($m_{3/2} \sim
  250~{\rm TeV}$) and a
  modulino with mass scale ($m_4 \sim 50~{\rm eV}$) within the FPF discovery reach.
\end{abstract}
  \maketitle
 
Neutrino oscillations imply the existence of new states that can
generate neutrino mass terms consistent with the Standard Model (SM)
$SU(2)$ gauge symmetry. Since the observed neutrino mass splittings
are tiny, a compelling realization comes from string models with large
extra dimensions~\cite{Antoniadis:1998ig}, in which gravitons and right-handed neutrinos are allowed to
propagate in the bulk, whereas the SM fields are
confined to localized 3-branes. Within this set up the 4-dimensional neutrino
Yukawa couplings are suppressed relative to charged-fermion Yukawa
couplings by a factor proportional to the square root of the volume of the extra
dimensions~\cite{Dienes:1998sb,Arkani-Hamed:1998wuz,Dvali:1999cn,Davoudiasl:2002fq,Antoniadis:2002qm}. In other words, neutrinos are very light for the
same reason that gravity appears to be very weakly coupled.

In addition, it has long been suspected that Planck
  ($M_p \simeq 2.4 \times 10^{18}~{\rm GeV}$) suppressed interactions could
be connected to neutrino physics, because the coupling $M_p^{-1} LLHH$
generates a mass $\langle H \rangle^2/M_p \sim 10^{-1}~{\rm meV}$ near
the neutrino mass scale,
where $L$ is the leptonic doublet and $H$ the Higgs doublet of the SM~\cite{Barbieri:1979hc,Akhmedov:1992hh}. It has also
been suspected that in superstring theory the superpartners of moduli
fields provide compelling candidates for the right-handed
neutrinos~\cite{Lukas:2000wn,Lukas:2000rg}. Before supersymmetry
(SUSY) breaking, some of these
moduli fields as well as their fermionic partners are exactly
massless. Masses for the moduli and modulinos are then generated 
when SUSY is broken and can be small. This may account for the
lightness of any sterile ($s$) neutrino states.

In this Letter we reexamine these captivating  ideas within the context of the swampland program~\cite{Vafa:2005ui},
focussing attention on neutrino-modulino oscillations which can be
probed by experiments at the CERN's Forward Physics Facility
(FPF)~\cite{Anchordoqui:2021ghd,Feng:2022inv}. We formulate a specific
scenario of high-scale SUSY breaking that can host a rather heavy gravitino and a
  modulino with mass scale within the FPF discovery reach. In a companion paper we discuss theoretical aspects of the model and explore in more detail the phenomenology of neutrino-modulino mixing in models with high- and low-scale
SUSY breaking~\cite{inpreparation}.

The swampland program keeps within bounds the set of 4-dimensional
effective field theories (EFTs) that are a low energy limit of quantum
gravity, and differentiate these theories from those that are not. The
plan of action is accomplished by conjecturing guiding principles that any EFT should
satisfy in order to be in the landscape of supersting theory vacua,
rather than be relegated to the swampland~\cite{Palti:2019pca,vanBeest:2021lhn,Agmon:2022thq}. Recently, by putting together predictions from the swampland program with
observational data it was elucidated that the smallness of
the cosmological constant in Planck units
($\Lambda \sim 10^{-120}M_p^4$) seems to indicate that our universe could stretch off in an asymptotic region of the string landscape of
vacua~\cite{Montero:2022prj}.

More concretely, the distance conjecture predicts the emergence
of infinite towers of Kaluza-Klein (KK) modes that become
exponentially light in Planck units ($M_p=1$), yielding a breakdown of any EFT at
infinite distance limits in moduli
space~\cite{Ooguri:2006in}. The related anti-de Sitter
(AdS) distance conjecture correlates the dark energy density to the
mass scale of the infinite tower of states~\cite{Lust:2019zwm}. Now, if the AdS
distance conjecture is to be generalized to dS space, the dark energy scale $\Lambda$
can be accommodated with the addition a mesoscopic (dark) extra-dimension characterized by a length-scale in the micron
range.

All in all, in this realization of a universe with 
tiny vacuum energy a dark dimension opens up at the characteristic mass scale of the KK tower,
\begin{equation}
m_{{\rm KK}_1} \sim \lambda^{-1} \ \Lambda^{1/4} \,,
\end{equation}
where $10^{-2} \alt \lambda \alt
10^{-4}$~\cite{Montero:2022prj}. The
5-dimensional Planck scale (or species
scale where gravity becomes strong) is
given by
\begin{equation}
 \Lambda_{\rm QG} \sim \sqrt{\frac{8\pi}{N}} \ M_p \sim \sqrt{8\pi} \
 \lambda^{-1/3} \ \Lambda^{1/12} \ M_p^{2/3} \,, 
\end{equation}
where $N$ is the number of the quantum field species below
$\Lambda_{\rm QG}$~\cite{Dvali:2007hz,Dvali:2007wp}, here to be identified with the number of KK modes. 

In the spirit of~\cite{Dienes:1998sb,Arkani-Hamed:1998wuz,Dvali:1999cn,Davoudiasl:2002fq,Antoniadis:2002qm}, we assume the generation of neutrino masses
originates in 5-dimensional bulk-brane interactions of the form
\begin{equation} 
  \mathscr{L}  \supset h_{ij} \ \overline L_i \ \tilde{H} \ \Psi_j(y=0) \,,
\end{equation}
where $\tilde{H} = -i\sigma_{2}H^{*}$, $L_i$ denotes the lepton
doublets (localized on the SM brane), $\Psi_j$ stands for the 3 bulk (right-handed) $R$-neutrinos
evaluated at the position of the SM brane, $y=0$ in the
fifth-dimension coordinate $y$, and $h_{ij}$ are coupling
constants. This gives a coupling with the
$L$-neutrinos of the form $\langle H \rangle \  \overline{\nu}_{L_i} \
\Psi_j (y=0)$, where $\langle H \rangle = 175~{\rm GeV}$ is the Higgs vacuum
expectation value. Expanding $\Psi_j$ into modes canonically normalized leads for each of them to a Yukawa $3 \times 3$ matrix suppressed by the square root of the volume of the bulk
$\sqrt{\pi R_\perp M_s}$, i.e.,
\begin{equation}
Y_{ij}= \frac{h_{ij}}{\sqrt{\pi R_\perp M_s}} \sim h_{ij} \frac{M_s}{M_p} \,,
\end{equation}
where $R_\perp \sim m_{{\rm KK}_1}^{-1}$ is the size of the dark
dimension and $M_s \alt \Lambda_{\rm QG}$ the string scale, and where
in the second rendition we have dropped factors of $\pi$'s and of the string coupling.

Now, neutrino oscillation data can be well-fitted in terms of two
nonzero differences $\Delta m^2_{ij} = m^2_i - m^2_j$ between the
squares of the masses of the three mass eigenstates; namely,
$\Delta m_{21}^2 =(7.53 \pm 0.18) \times 10^{-5}~{\rm eV}^2$ and
$\Delta m^2_{32} = (2.453 \pm 0.033) \times 10^{-3}~{\rm eV}^2$ or
$\Delta m^2_{32} = -(2.536 \pm 0.034) \times 10^{-3}~{\rm
  eV}^2$~\cite{ParticleDataGroup:2022pth}. It is easily seen that to
obtain the correct order of magnitude of neutrino masses the coupling
$h_{ij}$ should be of order $10^{-4}$ to $10^{-5}$ for
$10^9 \alt M_s/{\rm GeV} \alt 10^{10}$.

Note that KK modes of the
5-dimensional right-handed neutrino fields behave as an infinite tower
of sterile neutrinos, with masses proportional to $m_{{\rm
    KK}_1}$. However, only the lower mass states of the tower mix with
the active SM neutrinos in a pertinent fashion. The non-observation of
neutrino disappearance from oscillations into sterile neutrinos at
long- and short-baseline experiments places a 90\% CL upper limit on
the size of the dark dimension: $R_\perp < 0.4~\mu{\rm m}$ for the
normal neutrino ordering, and $R_\perp < 0.2~\mu{\rm m}$ for the
inverted neutrino
ordering~\cite{Machado:2011jt,Forero:2022skg}.\footnote{We arrived at
  these upper bounds by looking at the low mass limit of the lightest
  neutrino state in Fig.~6 of~\cite{Forero:2022skg} and rounding the numbers to one
  significant figure.} This set of parameters corresponds to $\lambda \alt 10^{-3}$ and so
$m_{{\rm KK}_1} \agt 2.5~{\rm eV}$~\cite{Anchordoqui:2022svl}.

Before proceeding, it is important to stress that the upper bounds
 on $R_\perp$ discussed in the previous paragraph are sensitive to assumptions of the $5^{\rm
  th}$ dimension geometry. Moreover,  in the presence of bulk masses,
the mixing of the first KK modes to active neutrinos can be
suppressed, and therefore the aforementioned bounds on $R_\perp$ can
be avoided~\cite{Carena:2017qhd,Anchordoqui:2023wkm}. It is also worth mentioning that such bulk masses have
the potential to increase the relative importance of the higher KK
modes, yielding distinct oscillation signatures via neutrino
disappearance/appearance effects. We
will discuss this multi-parameter scenario in~\cite{inpreparation}, herein we focus on the simplest
 one-parameter model unescorted by bulk masses.

Without further ado we bring into play the modulino. Following~\cite{Benakli:1997iu}, we assume that among the modulinos there is at
least one, $s$, with the following properties:
\begin{itemize}[noitemsep,topsep=0pt]
\item $s$ has only Planck mass suppressed interactions with SM fields, as expected for geometrical moduli governing the different couplings between light fields. We work within a simple construct, in which  the relevant light scale  for SM singlets is the gravitino mass $m_{3/2}$ and a dimensionless coupling constant with visible matter given by 
\begin{equation}
\lambda_i =\alpha_i \frac{ \ m_{3/2}}{M_p}, 
\label{lambda_i}
\end{equation}
 where $\alpha _i= {\cal O}(1)$.
 \item The mixing of $s$ with the active
SM neutrinos involves the electroweak symmetry breaking,
and the simplest appropriate effective operator is $\lambda_i \overline L_i
s H$. This operator generates a neutrino mass term $m_{\nu_i s} \bar \nu_i
s$ with  
\begin{equation}
m_{\nu_i s} = \alpha_i  \ \frac{m_{3/2} \langle H \rangle}{M_p} \, .
\label{mnu}
\end{equation}
\item The mass of the modulino, $m_{4}$, is induced via
  SUSY breaking. We assume that $m_4$ is absent at the level
  $m_{3/2}$ and appears as
  \begin{equation}
    m_4 = \beta \ \frac{m_{3/2}^2}{M_p} \,,
\label{m4}
  \end{equation}
where $\beta = {\cal O} (1)$.
\end{itemize}
In~\cite{inpreparation}, we will discuss these assumptions and the different realizations in details. Here, instead we will focus on the phenomenological implications.

To develop some sense for the orders of magnitude involved we make
contact with experiment. The FPF is a proposal to build a new underground
cavern at the Large Hadron Collider (LHC) to host a suite of
far-forward experiments during
the high-luminosity era~\cite{Anchordoqui:2021ghd,Feng:2022inv}. The
existing large LHC detectors have un-instrumented regions along the
beam line, and so miss the physics opportunities provided by the
enormous flux of particles produced in the far-forward direction. In
particular, the FPF proof of concept FASER has recently observed the
first neutrinos from LHC collisions~\cite{FASER:2023zcr}. During the
high-luminosity era, LHC collisions will provide an enormous flux of
neutrinos originating from the decay of light hadrons that can be used
to probe neutrino-modulino mixing. We particularize our calculations
to the design specifications of the FLArE detector, which will have a
$1~{\rm m} \times 1~{\rm m}$ cross sectional area with a 10~ton target
mass, and it will be located at about 620~m from the ATLAS interaction
point.

Considering typical energies ${\cal O} ({\rm TeV})$ of LHC neutrinos and a
baseline of $L \sim 620~{\rm m}$, FLArE will be sensitive to
modulino-neutrino square mass difference satisfying
\begin{equation}
  \frac{\Delta m^2_{41} L}{4E} = \frac{\pi}{2} \,,
\end{equation}
which implies $\Delta m^2_{41}
\sim 2000~{\rm eV}^2$~\cite{FASER:2019dxq}. It turns out that if
$m_{3/2} \sim 250~{\rm TeV}$, substituting for $\beta \sim
1.7$ into (\ref{m4}) we obtain the required $m_4$. We note in passing that the
abundance of unstable gravitinos with $m_{3/2} \alt 10~{\rm TeV}$ is severely constrained by the success of the big-bang nucleosynthesis~\cite{Kohri:2005wn}.

Now, we want to investigate whether SUSY models hosting 250~TeV
gravitinos live in the string landscape or the swampland. The first
step towards associating $m_{3/2}$ to the mass scale of an infinite
tower of KK modes was taken in~\cite{Antoniadis:1988jn}, and this
idea has been recently formulated as the gravitino conjecture~\cite{Cribiori:2021gbf,Castellano:2021yye}. The
dark dimension and the gravitino conjecture give rise to two
possible schemes, depending on the relation between the
corresponding towers of external states~\cite{Anchordoqui:2023oqm}. A first possibility is that $\Lambda$ and $m_{3/2}$ are connected to
the same KK tower. A second possibility is that the towers are
different.

In the single-tower scenario, the main formula of the gravitino
conjecture leads to 
\begin{equation}
  m_{3/2} = \left(\frac{\lambda_{3/2}}{\lambda}\right)^{1/n} \
  \left(\frac{\Lambda}{M_p^4}\right)^{1/(4n)} \ M_p \,,
\end{equation}
where $n$ and $\lambda_{3/2}$ are swampland parameters~\cite{Anchordoqui:2023oqm}. Note that for
fiducial values 
$\lambda \sim 10^{-4}$, $\lambda_{3/2} \sim 1$, and $n =2$ we obtain
$m_{3/2} \sim 250~{\rm TeV}$. This set of parameters lead to a
high-scale SUSY breaking
\begin{equation}
  M_{\rm SUSY} \sim \left(\frac{m_{3/2} \ M_p}{ \varkappa}\right)^{1/2}
  \sim 2.4 \times 10^{10}~{\rm GeV}\,, 
\end{equation}
where we have taken $\varkappa \sim 10^3$ to accommodate the requirement $M_{\rm SUSY} \alt
\Lambda_{\rm QG} \sim 2.6 \times 10^{10}~{\rm GeV}$.

Alternatively, we can assume that together with the infinite tower associated to the dark dimension there is a second tower of KK states related to $p$
 additional compact dimensions, with mass scale $m_{{\rm KK}_2}$. In this
case, the
quantum gravity cut-off is given by~\cite{Anchordoqui:2023oqm}
\begin{equation}
\Lambda_{\rm QG} = (8 \pi)^{2/(3+p)} \ m_{{\rm KK}_1}^{1/(3+p)} \ m_{{\rm
    KK}_2}^{p/(3+p)} \
M_p^{2/(3+p)} \,,  
\label{eqLamdaQG}
\end{equation}
and the gravitino conjecture yields
\begin{equation}
  m_{3/2} = \left(\lambda_{3/2} \frac{m_{{\rm
          KK}_2}}{M_p}\right)^{1/n} M_p \, .
\label{GSC}
\end{equation}
We select as fiducial parameters $\lambda \sim 10^{-4}$, $n=1$, $p=1$, $m_{{\rm KK}_2} \sim
10^9~{\rm GeV}$, and $\lambda_{3/2} \sim 2.5 \times 10^{-4}$. Our
choice leads to $m_{3/2} \sim 250~{\rm TeV}$ and $\Lambda_{\rm QG} \sim 1.4
\times 10^{10}~{\rm GeV}$. For the two-tower scheme the SUSY breaking
scale satisfies,
\begin{equation}
  M_{\rm SUSY} = \left(\frac{\lambda_{3/2} \ m_{{\rm
          KK}_2}}{\varkappa^n \
      M_p} \right)^{1/(2n)} M_p \,,
\label{msusy}
\end{equation}
  and so for $\varkappa \agt 10^{3.6}$ we have $M_{\rm SUSY} \alt
  \Lambda_{\rm QG}$.

Let's now provide a concise overview of concrete
realizations~\cite{inpreparation}. In linear supergravity realization,
the natural size expected for the gravitino mass is of order of the
compactification scale. A brane-localized gravitino mass term act as
deformation of the gravitino wave function in the bulk and simply
result in shifts in the KK mass spectrum by fractions of $\sim
1/{R_\perp}$. Working with only one flat extra-dimension leads
therefore to gravitino mass $\sim 1/{R_\perp}$ in the eV range, too
small for our purpose.   We instead contemplate the breaking of
supersymmetry through non-periodic boundary conditions or fluxes in
the other extra dimensions, anyway always present. To achieve $m_{3/2}
\sim \lambda_{3/2} m_{{\rm KK}_2} \sim 250~{\rm TeV}$ while employing a
dimension of size approximately $M_s^{-1}$, we must introduce a small
value for $ \lambda_{3/2}$. Within our framework, we can indeed consider the viability of a value 
around $\lambda_{3/2} \sim 10^{-4}$. However, in general questions 
regarding the compatibility of such a small parameter with the gravitino conjecture.
 Adopting $\lambda_{3/2}=1/2$, we need to consider this dimension to be significantly
larger than the string scale, as  $\Lambda_{\rm QG} \sim 2 \times
10^{9}~{\rm GeV}$ and $m_{{\rm KK}_2} \sim 5\times 10^5~{\rm
  GeV}$. 
  
  Alternatively, we can consider $p>1$ and invoke supersymmetry
breaking via fluxes, leading to
\begin{equation}
m_{3/2} = (\tilde\lambda_{3/2} \ m_{{\rm
    KK}_2})^{p} \ \frac{1}{M_s^{p-1}} \ .
\end{equation}
Then, by comparing with (\ref{GSC}) this corresponds to the choice 
$n=(p+3)/4p$ and $\lambda_{3/2}=\tilde\lambda_{3/2}^{p+3\over 4}(M_pR_\perp)^{p-1\over 4p}$.
 By setting $p=3$ and $\tilde\lambda_{3/2} =1/2$, we
obtain $m_{3/2} \sim 250~{\rm TeV}$. It addition, from 
(\ref{eqLamdaQG}) it follows that   
$M_s\sim \Lambda_{\rm QG} \sim 3  \times 10^{9}~{\rm GeV}$ and also that
$m_{{\rm KK}_2} \sim 2.6 \times 10^8~{\rm GeV}$. For a particular
example, we note that the modulino could be the fermionic partner of
the radion.\footnote{In the standard moduli stabilization by fluxes,
  all complex structure moduli and the dilaton are stabilized in a
  supersymmetric way while Kahler class moduli need an input from SUSY
  breaking. The radion is Kahler class and exists in a model
  independent fashion within the framework of the dark
  dimension.} In what follows, we concentrate
on this possible two-tower scheme and study the effects of
neutrino-modulino oscillations.

Before proceeding, we pause to note that the sharpened
  version of the weak gravity conjecture forbids the existence of
  non-SUSY AdS vacua supported by fluxes in a consistent quantum
  gravity theory~\cite{Ooguri:2016pdq}. Actually, the so-called non-SUSY AdS conjecture constrains the
  mass spectrum of the EFT in the far-infrared region, because the
  existence of AdS vacua depends on the neutrino masses and possible
  additional very light degrees of freedom via the Casimir potential~\cite{Arkani-Hamed:2007ryu}. In
  particular, the simplest EFT with 3 Majorana neutrinos would be
  rule out, because when the SM coupled to gravity is compactified down
  to three dimensions AdS vacua appear for any values of neutrino
  masses consistent with
  experiment~\cite{Ibanez:2017kvh}. For the model
  at hand, the Dirac neutrino states and associated KK modes prevent the rise of AdS vacua~\cite{Anchordoqui:2023wkm}. 

Next, in line with our stated plan, we show that neutrino-modulino oscillation effects can be
probed by neutrinos produced at the
LHC~\cite{Kling:2021gos}. Neutrino-modulino oscillations are analogous
to those in the 3+1 neutrino mixing framework, in which the flavor states
are given by the superposition of four massive neutrino states,
\begin{equation}
  |\nu_\alpha \rangle = \sum_{i =1}^4 U^*_{\alpha i} \ | \nu_i\rangle \,,
\end{equation}
where $\alpha = e,\mu, \tau, s$~\cite{Dasgupta:2021ies}. The presence of the modulino reshapes the active neutrino mixing angles via the unitarity
relations of the mixing matrix $\mathbb U$, i.e.,
\begin{equation}
\sum_{\alpha} U^*_{\alpha i} \ U_{\alpha j} = \delta_{ij} \quad \quad {\rm
  and} \quad \quad
  \sum_{i=1}^4 U^*_{\alpha i} \ U_{\beta i} = \delta_{\alpha \beta} \, ,
\label{uni2}
\end{equation}
where Greek indices run over the neutrino flavors and Roman indices
over the mass states.
Bearing this in mind, the unitary $4 \times 4$ mixing matrix takes the form 
$\mathbb U = \mathbb U_{34} \ \mathbb U_{24} \ \mathbb U_{14} \
\mathbb U_{\rm
  PMNS}$
where $\mathbb U_{\rm PMNS} = \mathbb U_{23} \,
\mathbb U_{13} \, \mathbb U_{12}$ is the the Pontecorvo-Maki-Nagakawa-Sakata (PMNS)
matrix~\cite{Pontecorvo:1967fh,Pontecorvo:1957qd,Maki:1962mu}.

Neutrino-modulino oscillations are driven by the Hamiltonian
\begin{equation}
  H = \frac{1}{2E} \ \mathbb U \ \mathbb M^2 \ \mathbb U^\dagger
\end{equation}
where $\mathbb M^2 \equiv {\rm diag} (0, \Delta m_{21}^2, \Delta m^2_{31}, \Delta
m^2_{41})$ and $E$ is the neutrino energy. The $\nu_{\alpha} \to \nu_{\beta}$ transition probability is found to be
\begin{equation}
  P_{\alpha \beta} = \left|\sum_{i=1}^4 = U^*_{\alpha
      i} \ U_{\beta
      i} \ \exp\left(-i \frac{\Delta m_{i1}^2 \ L}{2E} \right) \right|^2 \, ,
\end{equation}
where $L$ is the experiment baseline.  In the FPF-short-baseline limit
({\it viz.}, $\Delta m^2_{21}L/E \ll 1$ and $\Delta m^2_{31}L/E \ll 1$) for which SM oscillations have not
developed yet, the effective oscillation probabilities
can be written as 
\begin{equation}
 \!\!\!  P^{\rm FPF}_{\alpha \beta}  = \left|\sum_{i=1}^3 U^*_{\alpha i}
    U_{\beta i} + U^*_{\alpha 4}  U_{\beta 4}  \exp \left(-i
      \frac{\Delta m_{41}^2 L}{2E} \right)\right|^2 \! \! .
\label{tp1}  
\end{equation}
Using (\ref{uni2}), the transition probability (\ref{tp1}) can be recast as
\begin{eqnarray}
  P^{\rm FPF}_{\alpha \beta}  & = &  \left| \delta_{\alpha \beta} -
  U^*_{\alpha 4} \ U_{\beta 4} \left[ 1 - \exp\left(-i \frac{\Delta
                                                           m^2_{41} \ L}{2E} \right)\right]
                                                           \right|^2
                                                           \nonumber \\
& = &  \delta_{\alpha \beta} - \sin^2 2 \theta_{\alpha \beta} \ \sin^2  \left( \frac{\Delta m^2_{41}
     \  L}{4E} \right)  \, ,
\end{eqnarray}
where $\sin^2 2 \theta_{\alpha \beta} = 4 \ |U_{\alpha 4}|^2  \left(\delta_{\alpha
       \beta} - |U_{\beta 4}|^2\right)$.

\begin{figure}[tpb]
  \postscript{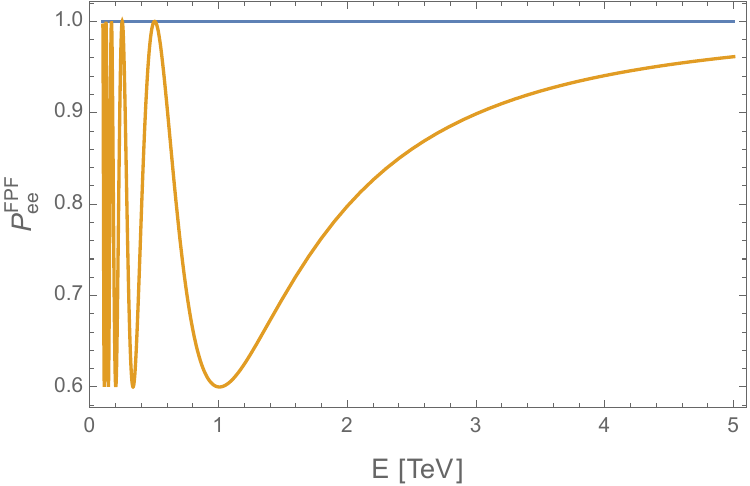}{0.9}
  \caption{The $\nu_e$ disappearance oscillation probability in the
    SM (blue) and in
    presence of a modulino with $\Delta m^2_{41}
    \sim 2000~{\rm eV}^2$,  $L = 620~{\rm m}$, and $|U_{e4}|^2 = 0.1$ (orange). \label{fig:1}}
\end{figure}  
\begin{figure}[tpb]
  \postscript{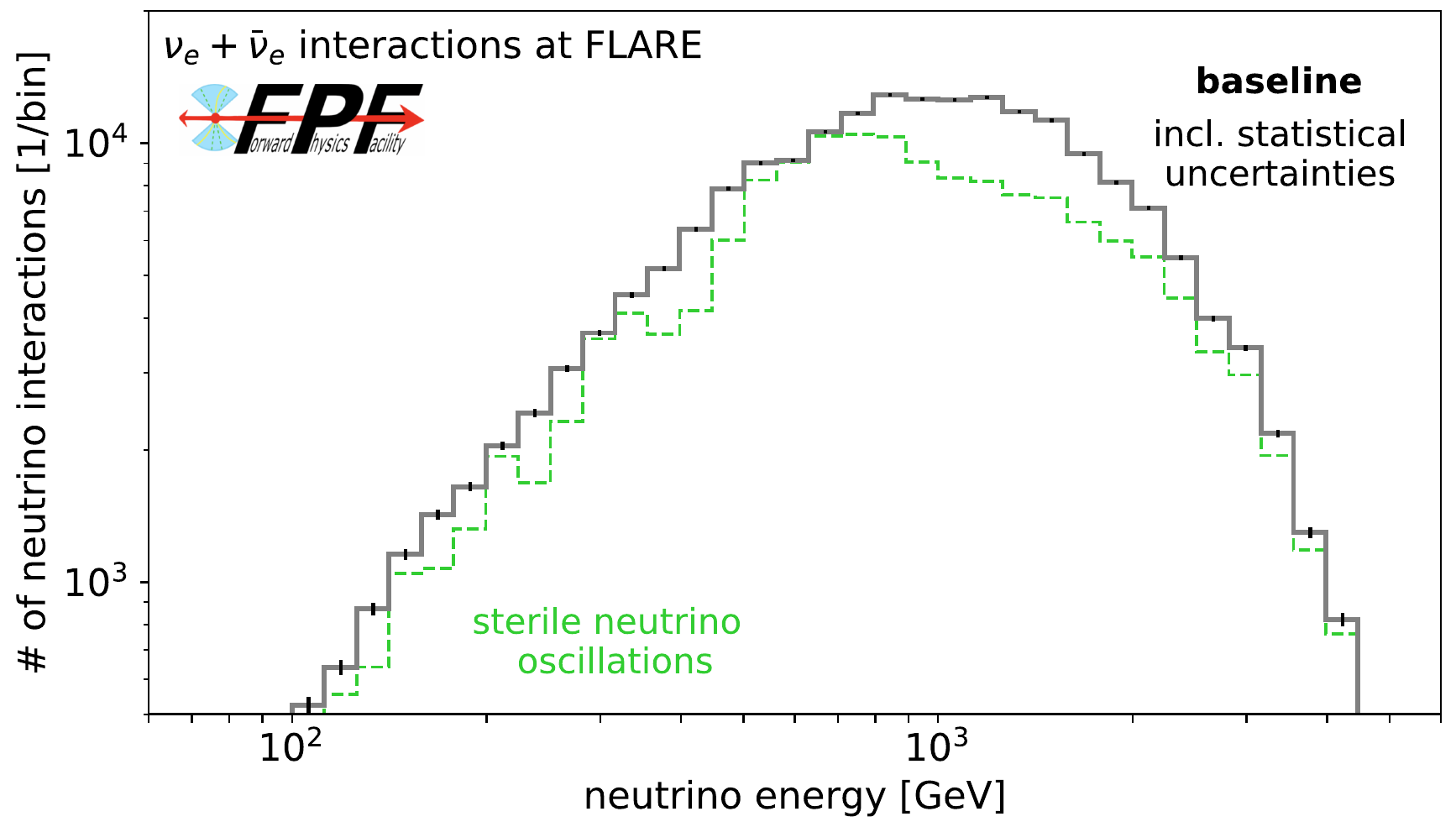}{0.9}
  \caption{The expected energy spectrum of interacting electron
    neutrinos in the FLArE detector is shown as the solid gray
    line. The corresponding statistical uncertainties are shown as
    black error bars. The colored dashed line shows how
    neutrino-modulino oscillations would change the expected flux if
the disappearance probability were characterized by $\Delta m_{41}^2 = 2500~{\rm eV}^2$ and
    $|U_{e4}|^2 =0.1$. The
    simulations have been normalized assuming an integrated luminosity
    of $3~{\rm ab}^{-1}$ and using the design specifications
    of the FLArE detector, which is supposed to have a $1~{\rm m}
    \times 1~{\rm m}$ cross sectional area, a 10~ton target mass, and
    $L = 620~{\rm m}$. Adapted from~\cite{Anchordoqui:2023P5}. \label{fig:2}}
\end{figure}

In Fig.~\ref{fig:1} we
show a comparison of the SM neutrino oscillation probability and the neutrino-modulino oscillation probability for the
disappearance channel assuming representative parameters. We can
see that using future FLArE data we will be able to constrain the 
neutrino-modulino mixing angle by
comparing the high-energy bins, where there is no oscillation of the
SM neutrinos, to the
bins at the neutrino-modulino oscillation minimum (around 1~TeV) where there is a deficit.

As an illustration, Fig.~\ref{fig:2} displays the
effect of electron neutrino disappearance due to a modulino of mass
$m_4 = 50~{\rm eV}$ with $|U_{e4}|^2 = 0.1$. It is clearly seen that the expected flux for the disappearance channel can
be distinguished from the prediction of the SM three
flavor scheme, demonstrating that the FPF could test the neutrino-modulino parameter space.

We end with a discussion of constraints on sterile
  neutrinos from astrophysical and cosmological observations, as well
  as searches in collider and beam
  dump experiments.
  \begin{itemize}[noitemsep,topsep=0pt]
\item At the intensity frontier sterile neutrinos
 can be produced abundantly via meson decay in beam-dump
 experiments. These experiments, however, probe very heavy sterile
 neutrinos with masses 
 in the range ${\rm MeV} \alt m_4  \alt {\rm GeV}$~\cite{Belle:2013ytx,NA62:2017qcd,Bryman:2019ssi,Bryman:2019bjg}. 
  \item Active-sterile
  neutrino mixing is strongly constrained for $m_4 \agt 100~{\rm keV}$
  to avoid excessive energy losses from supernova cores, but in the
  range $m_4 \alt 10~{\rm keV}$ the mixing angle is essentially
  unconstrained~\cite{Raffelt:2011nc}.
\item Big bang nucleosynthesis can also probe properties of sterile
  neutrinos. However, the $^2$H and $^4$He abundances  are mostly
  sensitive to the sterile neutrino lifetime and depends only weakly to the way the active-sterile mixing is distributed between flavors~\cite{Ruchayskiy:2012si}. 
\item Active neutrinos are produced through $\beta$-decays of unstable
  isotopes and in nuclear fission processes. Sterile neutrinos can
  also be produced via the active-sterile mixing if the sterile mass
  is smaller than the energy release of the relevant nuclear
  process. In this direction, the $\beta$-decay spectra of $^{18}$F, $^{19}$Ne, $^{35}$S, and $^{45}$Ca probe
  the range $4 < m_4/{\rm keV} <
  2000$~\cite{Calaprice:1983qn,Derbin:1993ni,Derbin:1997ut,Holzschuh:1999vy,Holzschuh:2000nj},
  whereas the spectra of the 
$^{187}$Re and $^3$H probe the mass range of interest to our investigation. An analysis of the $^{187}$Re beta decay leads
to the following 95\% CL upper limits: $|U_{e4}|^2 < 9
\times 10^{-3}$ at $m_4 = 1000~{\rm eV}$, $|U_{e4}|^2 < 1.2 \times 10^{-2}$ at $m_4
= 500~{\rm eV}$, $|U_{e4}|^2 < 4.4 \times 10^{-2}$ at $m_4= 200~{\rm eV}$, and $|U_{e4}|^2 < 0.116$ at
$m_4 = 100~{\rm eV}$~\cite{Galeazzi:2001py}. The analysis of tritium
$\beta$ decay data from the Mainz Neutrino Mass Experiment leads to
an 95\% CL upper bound $|U_{e4}|^2 \alt 0.1$ at $\Delta m_{41}^2 \sim
2500~{\rm eV}^2$~\cite{Kraus:2012he,Belesev:2012hx}, saturating the mixing strength
adopted for illustrative purposes in Fig.~\ref{fig:2}.  KATRIN Collaboration reported the most
restrictive bound from the tritium $\beta$-decay spectrum in the mass range $0.1 < m_4/{\rm keV} < 1.0$~\cite{KATRIN:2022spi}.
\end{itemize}
All in all, we conclude that there is a large window of sterile mixing
in the modulino mass range to be probed by experiments at the FPF. A
thorough study of the sensitivity of FPF experiments, including
systematic uncertainties, is beyond the scope of this paper. Actually, we
defer this type of study to the FPF Collaboration; first steps are
underway~\cite{Kling:2023tgr}.

In summary, we have put together a scenario with high-scale SUSY breaking
that can host a rather
heavy gravitino and a
modulino with a mass of about $50~{\rm eV}$. The corresponding models are
fully predictive through neutrino-modulino oscillations and can be
confronted with data to be collected by the FPF experiments.

\section*{Acknowledgments}

We thank Cumrun Vafa for discussion. The work of L.A.A. is supported by the U.S. National Science
Foundation (NSF Grant PHY-2112527). The work of D.L. is supported by the Origins
Excellence Cluster.

\end{document}